\documentclass{ws-mpla}
\usepackage{graphicx}

\def\d{\mbox{d}}
\def\lpder#1#2{\partial #1/\partial #2}

\def\Kds{Kerr--de~Sitter }

\def\oder#1#2{\frac{\d #1}{\d #2}}

\begin{document}

\markboth{Z. Stuchl\'{\i}k}
{Influence of the relict cosmological constant on accretion discs}

\catchline{}{}{}{}{}
%
%%%%%%%%%%%%%%%%%%%%%%%%%%%%%%%%%%%%%%%%%%%%%%%%%%%%%%%%%%%%%%%%%%%%

\title{INFLUENCE OF THE RELICT COSMOLOGICAL CONSTANT ON \\
       ACCRETION DISCS}

\author{\footnotesize ZDEN\v{E}K STUCHL\'{I}K}
  
\address{Institute of Physics, Faculty of Philosophy and Science, Silesian
         University in Opava \\
         Bezru\v{c}ovo n\'{a}m.~13, CZ-746\,01 Opava, Czech Republic \\
         Zdenek.Stuchlik@fpf.slu.cz}

\maketitle

\pub{Received (Day Month Year)}{Revised (Day Month Year)}

\begin{abstract}
Surprisingly, the relict cosmological constant has a crucial influence 
on properties of accretion discs orbiting black holes in quasars and active 
galactic nuclei. We show it by considering basic properties of both the 
geometrically thin and thick accretion discs in the Kerr--de Sitter black-hole 
(naked-singularity) spacetimes. Both thin and thick discs must have an outer 
edge allowing outflow of matter into the outer space, located nearby the so 
called static radius, where the gravitational attraction of a black hole 
is balanced by the cosmological repulsion. Jets produced by thick discs 
can be significantly collimated after crossing the static radius.
Extension of discs in quasars is comparable with extension of 
the associated galaxies, indicating a possibility that the relict cosmological 
constant puts an upper limit on extension of galaxies. 

\end{abstract}

\ccode{PACS Nos.: 04.25.-g, 04.70.Bw, 04.20.Dw, 98.62.Mw}

\section{Introduction}

Recent data from a wide variety of independent cosmological tests indicate 
convincingly that within the framework of the inflationary cosmology a  
non-zero, although very small, vacuum energy density, i.e., a relict 
repulsive cosmological constant (RRCC), $\Lambda > 0$, or some kind of
similarly  
acting quintessence, has to be invoked in order to explain the dynamics of the
recent Universe.\cite{Bah-etal:1999:SCIEN:,Kol-Tur:1990:EarUni:}

There is a strong ``concordance'' indication\cite{Spe-etal:2003:ASTJS:}
that the observed value of the vacuum energy density is
\begin{equation}
      \varrho_{\mathrm{vac(0)}}\approx 0.73 \varrho_{\rm crit(0)}
\end{equation}
with present values of the critical energy density
$\varrho_{\rm crit(0)}$, and the Hubble parameter $H_{0}$ given by
\begin{equation}
      \varrho_{\rm crit(0)}=\frac{3H_{0}^2}{8\pi},\quad H_{0}=100h\
      \rm{km}\ \rm{s}^{-1}\ \rm{Mpc}^{-1}.
\end{equation}
Taking value of the dimensionless parameter $h\approx 0.7$, we obtain
the RRCC to be
\begin{equation}
      \Lambda_{0}=8\pi\varrho_{\mathrm{vac(0)}}\approx 1.3 \times 10^{-56}\
      \rm{cm}^{-2}. 
\label{E3}
\end{equation}

It is well known that the RRCC strongly influences expansion of the Universe,
leading finally to an exponentially accelerated
stage.\cite{Car-Ost:1996:ModAst:}
However, surprisingly enough, the RRCC can be relevant for accretion processes
in the field of central black holes in quasars and active galactic nuclei. 

Basic properties of  
geometrically thin accretion discs with low accretion rates and negligible
pressure 
are given by the circular geodetical motion in the black-hole
backgrounds,\cite{Nov-Tho:1973:BlaHol:436} while for geometrically thick discs
with high 
accretion rates and relevant pressure they are determined by equipotential
surfaces of test perfect fluid rotating in the
backgrounds.\cite{Koz-Jar-Abr:1978:ASTRA:,Abr-Jar-Sik:1978:ASTRA:}
The presence of the RRCC changes substantially the asymptotic structure of the
black-hole 
(naked-singularity) backgrounds as they become asymptotically de Sitter and 
contain a cosmological event horizon behind which the spacetime is dynamic.
Properties of the circular geodesic orbits in the Schwarzschild--de~Sitter
(SdS) and Reissner--Nordstr\"{o}m--de~Sitter (RNdS) spacetimes show that 
due to the presence of the RRCC, the thin discs have not only an inner edge
determined (approximately) by the 
radius of the innermost stable circular orbit, but also an outer edge given by
the radius of the outermost stable circular orbit, located nearby the 
static radius.\cite{Stu-Hle:1999:PHYSR4:,Stu-Hle:2002:ACTPS2:}
The vicinity of the 
static radius can be considered as a counterpart to the asymptotically flat
region of the Kerr spacetimes, as can be demonstrated by the  
embedding diagrams of the equatorial plane of both the directly projected
geometry 
and the optical reference geometry reflecting some hidden properties of the
geodesic motion.\cite{Abr-Pra:1990:MONNR:,Stu-Hle:2000:CLAQG:,Hle:2002:JB60:}
The analysis of equilibrium configurations of perfect fluid orbiting in
the SdS black-hole backgrounds shows a possible existence of thick discs with
outflow of 
matter through an outer cusp of the equilibrium configuration due to violation
of
mechanical equilibrium.\cite{Stu-Sla-Hle:2000:ASTRA:} Such an outflow can 
represent a strong stabilizing effect\cite{Rez-Zan-Fon:2003:ASTRA:} against
the runaway instability\cite{Abr-Cal-Nob:1983:NATURE:} of thick discs orbiting
the SdS black holes.

However, it is crucial to understand the role of the RRCC in astrophysically
most relevant, rotating Kerr backgrounds. In the Kerr--de~Sitter (KdS)
backgrounds, we shall consider circular equatorial motion of test particles, 
relevant for thin discs, and equilibrium configurations of perfect fluid, 
relevant for thick discs. We shall focus attention on the black-hole
backgrounds, 
but some results related to the naked-singularity backgrounds will be
mentioned
because of increasing theoretical evidence on possible existence of naked
singularities.\cite{deFel-Yun:2001:CLAQG:} 

\section{Kerr--de~Sitter spacetimes}
\label{KdS spacetimes}

In the standard Boyer--Lindquist coordinates ($t,r,\theta, \phi$) and the
geometric units ($c=G=1$), the Kerr--(anti-)de Sitter geometry is given by the
line element
\pagebreak
\begin{eqnarray}
               \d s^2 =  & - & \frac{\Delta_r}{I^2 \rho^2}(\d t-a\sin^2 \theta
                \d\phi)^2
                + \frac{\Delta_{\theta}\sin^2 \theta}{I^2 \rho^2}
                \left[a\d t- \left(r^2+a^2 \right) \d\phi \right]^2 \nonumber\\
                         & + & \frac{\rho^2}{\Delta_r} \d r^2 +
                \frac{\rho^2}{\Delta_{\theta}} \d\theta^2,  
                                                               \label{e1}
\end{eqnarray}
where
\begin{eqnarray}
        \Delta_r & = & -\frac{1}{3}\Lambda r^2 \left(r^2+a^2 \right)
                +r^2 -2Mr + a^2,  \\                            \label{e2}
        \Delta_{\theta} & = & 1+ \frac{1}{3} \Lambda a^2 \cos^2 \theta, 
  \quad I = 1 + \frac{1}{3} \Lambda a^2, 
  \quad \rho^2 = r^2 +a^2 \cos^2 \theta.                    \label{e3}
\end{eqnarray}
The parameters of the spacetime are: mass ($M$), specific angular
momentum ($a$), cosmological constant ($\Lambda$). It is convenient to
introduce a 
dimensionless cosmological parameter
\begin{equation}                                                \label{e6}
        y = \frac{1}{3} \Lambda M^2.
\end{equation}
For simplicity, we put $M=1$ hereafter. Equivalently, also the coordinates
$t,\ r$, the line element $\d s$, and the rotational parameter of the
spacetime $a$, being 
expressed in units of $M$, become dimensionless.
We focus our attention to the case $y>0$ corresponding to the repulsive
cosmological constant; then (\ref{e1}) describes a KdS spacetime.

The event horizons of the spacetime are given by the pseudosingularities of
the line element (\ref{e1}), determined by the condition $\Delta_r =0$. The
loci of the event horizons are implicitly determined by the relation
\begin{equation}                                                   \label{e7}
        a^2 = a^2_{\mathrm{h}}(r;y) \equiv \frac{r^2 -2r -yr^4}{yr^2 -1}.
\end{equation}
It can be shown\cite{Stu-Sla:2004:PHYSR4:} that a critical value
of the cosmological parameter exists
\begin{equation}                                                   \label{e11}
        y_{\mathrm{c(KdS)}} = \frac{16}{(3+2\sqrt{3})^3} \doteq 0,05924,
\end{equation}
such that for $y>y_{\mathrm{c(KdS)}}$, only naked-singularity backgrounds
exist for $a^2>0$. There is another critical value              
$y_{\mathrm{c(SdS)}} = 1/27 \doteq 0.03704$,
which is limiting the existence of SdS black holes.\cite{Stu-Hle:1999:PHYSR4:}
In the RNdS spacetimes, the critical value is\cite{Stu-Hle:2002:ACTPS2:}
$y_{\mathrm{c(RNdS)}} = 2/27 \doteq 0.07407$.

If $y = y_{\mathrm{c(KdS)}}$, the function $a^2_{\mathrm{h}}(r;y)$ has an
inflex point corresponding to a critical value of the
rotation parameter of the KdS spacetimes
\begin{equation}                                                  \label{e13}
        a^2_{\rm crit} = \frac{3}{16} (3+ 2\sqrt{3}) \doteq 1,21202.
\end{equation}
KdS black holes can exist for $a^2 < a^2_{\rm crit}$ only,
while KdS naked singularities can exist for both $a^2 <
a^2_{\rm crit}$ and $a^2 > a^2_{\rm crit}$.

\begin{figure}
\centering
\includegraphics[width=.6\hsize]{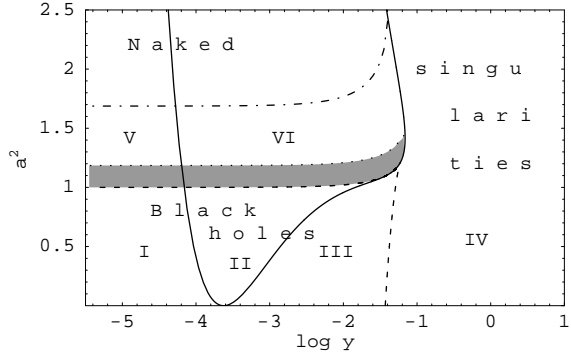}
\caption{Classification of the Kerr--de~Sitter spacetimes. Dashed curves
  separate black holes and naked singularities. Full curves divide the
parametric space by properties of the stable circular orbits relevant  
for Keplerian accretion discs. Spacetimes with both plus-family and
minus-family stable circular orbits ({\bf I} and {\bf V}). Spacetimes with no
minus-family stable circular orbits ({\bf II} and {\bf VI}). 
Spacetimes with no stable circular orbits ({\bf III} and {\bf IV}).
Dashed-dotted curve
defines the subregion of the naked-singularity spacetimes, where the 
plus-family circular orbits could be stable and counter-rotating (from the
point of view of the LNRF), shaded is the subregion
allowing stable circular orbits with $E_{+}<0$! (Taken
from Ref.~\protect\refcite{Stu-Sla:2004:PHYSR4:}.)}
\label{f1}
\end{figure}

Separation of the KdS black-hole and
naked-singularity spacetimes in the parameter space $y$--$a^2$ is shown in
Fig.~\ref{f1}.  In the black-hole spacetimes there are two black-hole
horizons and the cosmological horizon, with $r_{\mathrm{h}-} < r_{\mathrm{h}+}
< r_{\mathrm{c}}$. In the naked-singularity spacetimes, there is the
cosmological horizon $r_{\mathrm{c}}$ only.

%%%%%%%%%%%%%%%%%%%%%%%%% figure 1 %%%%%%%%%%%%%%%%%%%%%%%%%%%%%%%%%%%%%%%%%%%
%%%%%%%%%%%%%%%%%%%%%%%%%%%%%%%%%%%%%%%%%%%%%%%%%%%%%%%%%%%%%%%%%%%%%%%%%%%%%%%

The extreme cases, when two (or all three) horizons coalesce, were discussed
in detail for the case of RNdS
spacetimes.\cite{Bri-Hay:1994:CLAQG:,Hay-Nak:1994:PHYSR4:}  
In the KdS spacetimes, the situation is analogical. If
$r_{\mathrm{h}-}=r_{\mathrm{h}+} < 
r_{\mathrm{c}}$, the extreme black-hole case occurs, if $r_{\mathrm{h}-} <
r_{\mathrm{h}+} = r_{\mathrm{c}}$, the marginal naked-singularity case occurs,
if $r_{\mathrm{h}-} = r_{\mathrm{h}+} = r_{\mathrm{c}}$, the ``ultra-extreme''
case occurs corresponding to a naked~singularity.

\section{Thin discs}
\label{Thin}

Basic properties of thin accretion discs are determined by equatorial circular
motion of test particles because any
tilted disc has to be driven to the equatorial plane of the rotating
spacetimes due to the dragging of inertial frames.\cite{Bar-Pet:1975:ASTRJ2L:}

The motion of a test particle with rest mass $m$ is given by the geodesic
equations. In a separated and integrated form,  the equations were obtained
by Carter.\cite{Car:1973:BlaHol:} For the motion restricted to the equatorial 
plane ($\d\theta/\d \lambda = 0$, $\theta = \pi/2$) of the KdS spacetime, 
the Carter equations take the following form
\begin{eqnarray}
        r^2 \oder{r}{\lambda} & = & \pm R^{1/2} (r), \\
                                                                \label{e14}
        r^2 \oder{\phi}{\lambda} & = & - IP_{\theta}+
                \frac{a I P_r}{\Delta_r}, \\
                                                                \label{e15}
        r^2 \oder{t}{\lambda} & = & -a I P_{\theta} +
                \frac{(r^2 +a^2) I P_r}{\Delta_r},
                                                                \label{e16}
\end{eqnarray}
where
\begin{eqnarray}
     & & R(r) = P^2_r -\Delta_r \left(m^2r^2 +K \right),  \\
                                                                \label{e17}
     & & P_r = I {\cal{E}} \left(r^2 +a^2 \right) - I a \Phi, 
  \quad P_{\theta} = I (a {\cal{E}} - \Phi),                            
  \quad K = I^2 (a {\cal{E}} - \Phi)^2.                         \label{e18}
\end{eqnarray}
The proper time of the particle $\tau$ is related to the affine parameter
$\lambda$ by $\tau = m \lambda$. The constants of motion are: energy
($\cal{E}$), related to the stationarity of the geometry, axial angular
momentum ($\Phi$), related to the axial symmetry of the geometry, `total'
angular momentum ($K$), related to the hidden symmetry of the geometry. For
the equatorial motion, $K$ is restricted through Eq.~(\ref{e18}) following
from the conditions on the latitudinal motion.\cite{Stu:1983:BULAI:} Notice
that $\cal{E}$ and $\Phi$ cannot be interpreted as energy and axial angular
momentum at infinity, since the spacetime is not asymptotically flat.

The equatorial motion is governed by the constants of motion
${\cal{E}},\ \Phi$. Its properties can
be conveniently determined by an ``effective potential'' given by the
condition $R(r) = 0$ for turning points of the radial motion. It is useful
to define specific energy and specific angular momentum by the relations
\begin{equation}                                                \label{e21}
        E \equiv \frac{I {\cal{E}}}{m}, \ L \equiv \frac{I \Phi}{m}.
\end{equation}
Solving the equation $R(r) = 0$,
we find the effective potential in the form
\begin{eqnarray}
        E_{(\pm)}(r;L,a,y) &\equiv&
\left[\left(1+y a^2 \right)r\left(r^2 +a^2 \right)
  + 2a^2 \right]^{-1}\nonumber\\
&\times&
\bigg\{a\left[yr\left(r^2+a^2 \right)+2
                \right]L\nonumber\\
&\pm& \left.\Delta^{1/2}_r
                \left\{r^2 L^2 +r\left[\left(1+ya^2\right)r\left(r^2+a^2
                \right)+ 2a^2 \right] \right\}^{1/2}\right\}.   \label{e23}
\end{eqnarray}
In the stationary regions ($\Delta_r \geq 0$), the motion is allowed
where\cite{Bic-Stu-Bal:1989:BULAI:}
\begin{equation}                                                    \label{e24}
        E \geq E_{(+)}(r;L,a,y).
\end{equation}

The equatorial circular orbits can be determined by
solving simultaneously the equations $R(r) = 0, dR/dr = 0$.
The specific energy of the orbits is given by 
\begin{equation}                                                 \label{e38}
        E_{\pm}(r;a,y) = \frac{1-\frac{2}{r}- \left(r^2+ a^2 \right)y
                \pm a \left(\frac{1}{r^3}- y\right)^{1/2}}
                {\left[1- \frac{3}{r}- a^2y\pm 2a \left(\frac{1}{r^3}- y
                \right)^{1/2} \right]^{1/2}},
\end{equation}
while the specific angular momentum of the orbits is determined by
\begin{equation}                                                     
       L_{\pm}(r;a,y) =
                - \frac{2a +ar\left(r^2 +a^2 \right)y
                \mp r\left(r^2 +a^2\right)
                \left(\frac{1}{r^3} -y \right)^{1/2}}
                {r \left[1 -\frac{3}{r}- a^2y\pm 2a
                \left(\frac{1}{r^3} -y\right)^{1/2}
                \right]^{1/2}}.     
\label{e40}
\end{equation}
The relations (\ref{e38})--(\ref{e40}) determine two families of the circular
orbits. We call them plus-family orbits and minus-family
orbits\cite{Stu-Sla:2004:PHYSR4:}
according to the $\pm$ sign in the relations (\ref{e38})--(\ref{e40}).
Inspecting expressions (\ref{e38}) and (\ref{e40}), we find two reality
restrictions on the circular orbits. The first one is given by the relation
\begin{equation}                                              \label{e49}
        y \leq y_{\mathrm{s}} \equiv \frac{1}{r^3},
\end{equation}
which introduces the notion of the ``static radius'', given by the formula
$r_{\mathrm{s}} = y^{-1/3}$ independently of the rotational parameter $a$. It
can be compared with formally identical result in the Schwarzschild--de~Sitter
spacetimes.\cite{Stu-Hle:1999:PHYSR4:} A ``free'' or ``geodetical'' observer
on the static radius has only $U^t$ component of 4-velocity being non-zero. The
position on the static radius is unstable relative to radial perturbations.
The second restriction is given by the condition 
\begin{equation}                                                \label{e50}
        1 -\frac{3}{r} -a^2y \pm 2a\left(\frac{1}{r^3}
                -y \right)^{1/2} \geq 0;
\end{equation}
the equality determines photon circular orbits with $E \to \infty$ and $L \to
\pm \infty$. 
The photon circular orbits of the plus-family are given by the relation
\begin{equation}                                               \label{e51}
  a = a^{(+)}_{\mathrm{ph(1,2)}}(r;y) \equiv \frac{\left(1 -yr^3\right)^{1/2}
                \pm \left(1 -3yr^2 \right)^{1/2}}
                {yr^{3/2}},
\end{equation}
while for the minus-family orbits they are given by
\begin{equation}                                              \label{e52}
  a = a^{(-)}_{\mathrm{ph(1,2)}}(r;y) \equiv \frac{-\left(1 -yr^3 \right)^{1/2}
                \pm \left(1 -3yr^2 \right)^{1/2}}
                {yr^{3/2}}.
\end{equation}
A detailed discussion of the photon circular orbits can be found
in Refs.~\refcite{Stu-Hle:2000:CLAQG:,Stu-Sla:2004:PHYSR4:}.

The behaviour of circular orbits in the field of Kerr black holes ($y=0$)
suggests that the plus-family orbits correspond to the co-rotating orbits,
while the minus-family circular orbits correspond to the counter-rotating
ones.  However, this statement is not correct even for Kerr naked-singularity 
spacetimes with the rotational parameter low enough, where counter-rotating 
plus-family orbits could exist nearby the ring
singularity.\cite{Stu:1980:BULAI:}  
In the KdS spacetimes we cannot
identify the plus-family circular orbits with purely co-rotating orbits even
in the black-hole spacetimes; moreover, it is not possible to define the
co-rotating 
(counter-rotating) orbits in relation to stationary observers at infinity, as
can be done in the Kerr spacetimes, since the KdS spacetimes are not
asymptotically flat.

Orientation of the circular orbits in the KdS spacetimes must be related to 
locally non-rotating frames (LNRF), similarly to the case of asymptotically
flat Kerr spacetimes. In the KdS spacetimes, the tetrad
of 1-forms corresponding to the LNRF is given by
Ref.~\refcite{Stu-Hle:2000:CLAQG:}:
\newline
\begin{minipage}{.4\linewidth}
\begin{eqnarray}
\omega^{(t)}&\equiv&\left
(\frac{\Delta_{r}\Delta_{\theta}\varrho^{2}}{I^{2}A}\right )^{1/2}{\rm d}t,
\nonumber \\
\omega^{(r)}&\equiv&\left (\frac{\varrho^{2}}{\Delta_{r}}\right )^{1/2}{\rm
  d}r, \nonumber
\end{eqnarray}
\end{minipage}\hfill
\begin{minipage}{.58\linewidth}
\begin{eqnarray}
\omega^{(\phi)}&\equiv&\left (\frac{A\sin^{2}\theta}{I^{2}\varrho^{2}}\right
)^{1/2}({\rm d}\phi-\Omega {\rm d}t), \\ 
\omega^{(\theta)}&\equiv&\left (\frac{\varrho^{2}}{\Delta_{\theta}}\right
)^{1/2}{\rm d}\theta, 
\end{eqnarray}
\end{minipage}
\vskip2ex\noindent
with the angular velocity of the LNRF being given by
\begin{equation}
\Omega\equiv\frac{{\rm d}\phi}{{\rm d}t}=\frac{a}{A}\left
                 [-\Delta_{r}+(r^{2}+a^{2})\Delta_{\theta}\right ];\quad 
A\equiv (r^{2}+a^{2})^{2}-a^{2}\Delta_{r}.
\end{equation}

Locally measured components of 4-momentum in the LNRF are given by the
projection 
of a particle's 4-momentum  onto the tetrad
\begin{equation}
p^{(\alpha)}=p^{\mu}\omega^{(\alpha)}_{\mu}, \quad
p^{\mu}=m\frac{{\rm d}x^{\mu}}{{\rm d}\tau}\equiv m\dot{x}^{\mu}=\frac{{\rm
    d}x^{\mu}}{{\rm d}\lambda}.
\end{equation}
A simple calculation reveals the intuitively anticipated relation
\begin{equation}
p^{(\phi)}=\frac{mr}{A^{1/2}}L.
\end{equation}
We can see that the sign of the azimuthal component of the 4-momentum measured
in the LNRF is given by the sign of the specific
angular momentum of a particle on the orbit of interest. Therefore the
circular orbits with $p^{(\phi)}>0,\ (L>0)$, we call co-rotating, and the
circular orbits with $p^{(\phi)}<0,\ (L<0)$ we call counter-rotating, in
agreement with the case of asymptotically flat Kerr spacetimes.

The circular geodesics can be astrophysically relevant, if they are stable
with respect to radial perturbations. 
The loci of the stable circular orbits are given by the condition
\begin{equation}                                                     
        \frac{\d^2 R}{\d r^2} \geq 0
                                     \label{e56}
\end{equation}
that has to be satisfied simultaneously with the conditions $R(r) =0$ and
$\d R/\d r = 0$ determining the circular orbits. The radii of the stable orbits
of both families are restricted by the condition\cite{Stu-Sla:2004:PHYSR4:}
\begin{equation}                                                     
        r \left[6 -r +r^3 (4r -15)y \right] \mp
         8a \left[r\left(1 -yr^3 \right)^3\right]^{1/2} + 
                 a^2 \left[3 +r^2y\left(1 -4yr^3 \right)\right] \geq 0. 
                                                \label{e57}
\end{equation}
The marginally stable orbits of both families are described by the relation 
\begin{eqnarray}
  a^2 &=& a^2_{\mathrm{ms(1,2)}}(r;y) \equiv
    \left[3 +r^2y \left(1-4yr^3\right)\right]^{-2}r
    \bigg\{\left[r -6 -r^3 (4r -15)y \right]\nonumber \\
    &\times&
      \left[3 +r^2y \left(1 -4yr^3 \right)\right]
      +32 \left(1 -yr^3 \right)^3 \pm 8 \left(1-yr^3\right)^{3/2}
      \left(1 -4yr^3 \right)^{1/2}\nonumber \\
    &\times& \left.\left\{r \left[3 -ry \left(6 +10r -15yr^3\right) \right]
      -2 \right \}^{1/2}\right \}.    \label{e58}
\end{eqnarray}
The ($\pm$) sign in Eq.~(\ref{e58}) is not directly related to the
plus-family and the minus-family orbits. The function $a^2_{\mathrm{ms(1)}}$,
corresponding to the $(+)$ sign in Eq.~(\ref{e58}), determines marginally
stable orbits of the plus-family, while the function $a^2_{\mathrm{ms(2)}}$,
corresponding to the $(-)$ sign in Eq.~(\ref{e58}), is relevant for both the
plus-family and minus-family orbits. A detailed analysis shows that the
critical value of the cosmological parameter for the existence of the stable
(plus-family) orbits is given by
\begin{equation}
 y_{\mathrm{crit(ms+)}} = \frac{100}{(5 + 2\sqrt{10})^{3}}\doteq 0.06886.
                                                        \label{ee1}
\end{equation}
No stable circular orbits (of any family) exist for $y>y_{\mathrm{crit(ms+)}}$.
The critical value of $y$ for the existence of the minus-family stable
circular orbits is given by
\begin{equation}
          y_{\mathrm{crit(ms-)}} = \frac{12}{15^{4}}.             \label{ee3}
\end{equation}
It coincides with the limit on the existence of the stable circular orbits in
the SdS spacetimes.\cite{Stu-Hle:1999:PHYSR4:} In the parameter space
$y$--$a^{2}$, separation of the KdS spacetimes according to the
existence of stable circular orbits is given in Fig.~\ref{f1}.

\begin{figure}
\centering
\includegraphics[width=.6\hsize]{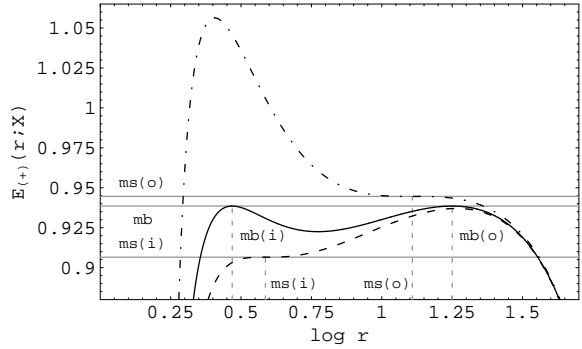}
\caption{Effective potentials of equatorial radial motion of test particles in
  \Kds black-hole spacetime ($y=10^{-4},\ a=0.6$)
allowing stable circular orbits for co-rotating particles. Marginally bound
(mb) orbits are given by the solid curve corresponding to $X=X_{mb+}\doteq
2.38445$. The curve has two local maxima of the same
value, $E_{mb}\doteq 0.93856$, leading to inner (mb(i)) and outer (mb(o))
marginally bound orbits. The dashed effective
potential defines inner marginally stable orbit (ms(i)) by coalescing local
minimum and (inner) local maximum, and corresponds to $X=X_{ms(i)+}\doteq
2.20307$ with energy $E_{ms(i)+}\doteq 0.90654$. 
In an analogous manner the dashed-dotted potential defines 
outer marginally stable orbit (ms(o)) with energy $E_{ms(o)+}\doteq 0.94451$
corresponding to $X=X_{ms(o)+}\doteq 2.90538$. (Taken from
Ref.~\protect\refcite{Stu-Sla:2004:PHYSR4:}.)}
\label{f2}
\end{figure}

Behaviour of the effective potential (\ref{e23}) enables us to introduce the
notion of marginally bound orbits, i.e., unstable circular orbits where a
small radial perturbation causes infall of a particle from the orbit to the
centre, or its escape to the cosmological horizon. For some special value of
the axial parameter $X = L - aE$, denoted as $X_{\mathrm{mb}}$, the effective
potential has two local maxima related by 
\begin{equation}
   E_{(+)}(r_{\mathrm{mb(i)}};X_{\mathrm{mb}},a,y)
     =E_{(+)}(r_{\mathrm{mb(o)}};X_{\mathrm{mb}},a,y),
\end{equation}
corresponding to both the inner and outer marginally bound orbits, see
Fig.~\ref{f2}.  
For completeness, the figure includes the effective
potentials defining both the inner and outer marginally stable orbits.  
The search for the marginally bound orbits in a concrete KdS spacetime
admitting stable circular orbits must be realized in a numerical way. 
Clearly, in the spacetimes with
$y \geq 12/15^4$, the minus-family marginally bound orbits do not
exist. In the spacetimes admitting stable plus-family orbits, there is
$r_{\rm mb(o)} \sim r_{\rm s}$ but $r_{\rm ms(o)} \sim 0.7 r_{\rm s}$.  

%%%%%%%%%%%%%%%%%%%%%%%%%%% figure 2 %%%%%%%%%%%%%%%%%%%%%%%%%%%%%%%%%%%%%%%
%%%%%%%%%%%%%%%%%%%%%%%%%%%%%%%%%%%%%%%%%%%%%%%%%%%%%%%%%%%%%%%%%%%%%%%%%%%%

In comparison with the asymptotically flat Kerr spacetimes, where the effect
of spacetime rotation vanishes for asymptotically large values of the radius,
in the KdS spacetimes the properties of the circular orbits must
be treated more carefully, because the rotational effect is relevant in the
whole region where the circular orbits are allowed and it survives even at the
cosmological horizon.

The minus-family orbits have $L_-<0$ in
each KdS spacetime and such orbits are counter-rotating relative to the LNRF. 

In the black-hole spacetimes, the plus-family orbits are co-rotating in almost
all radii where the circular orbits are allowed except some region in vicinity
of the static radius, where they become to be counter-rotating. However, these
orbits are unstable. 
The specific angular momentum and energy of particles located at the static
radius, where 
the plus-family and minus-family orbits coalesce, are given by 
\begin{equation}
L_{\mathrm{s}} = -a\frac{3y^{1/3}+a^{2}y}{\left (1-3y^{1/3}-a^{2}y
  \right )^{1/2}}, \quad
E_{\mathrm{s}} = (1-3y^{1/3}-a^2 y)^{1/2}.
\end{equation}

In the naked-singularity spacetimes, the plus-family
orbits behave in a more complex way. They are always
counter-rotating in vicinity of the static radius. Moreover, in the naked
singularity spacetimes with the rotational parameter low enough
($a<3\sqrt{3}/4,\ y=0$),   
stable counter-rotating plus-family circular orbits exist.
When $a$ is very close to the extreme hole state ($a<4\sqrt{2}/(3\sqrt{3}),\
y=0$), even stable plus-family orbits with $E < 0$ can
exist,\cite{Stu-Sla:2004:PHYSR4:} see Fig.~\ref{f1}.

Angular velocity $\Omega=\d\phi/\d t$ of a thin, Keplerian accretion disc is
given by
\begin{equation}                                                     \label{e5}
     \Omega_{\rm K\pm} = \pm\frac{1}{r^{3/2}/(1-yr^3)^{1/2} \pm a}.
\end{equation}
Matter in the thin disc spirals from the outer marginally stable orbit
through the sequence of stable circular orbits down to the inner marginally
stable orbit losing energy and angular momentum due to the viscosity. The
necessary conditions for such a differential rotation
\begin{equation}                                                     \label{E6}
     \frac{\d\Omega_{\rm K+}}{\d r}<0 \quad \frac{\d L_+}{\d r}\geq 0 
     \qquad\mbox{or}\qquad 
     \frac{\d\Omega_{\rm K-}}{\d r}>0 \quad \frac{\d L_-}{\d r}\leq 0, 
\end{equation}
are fulfilled by the relations (\ref{e40}) and (\ref{e5}). The efficiency of
accretion, i.e., the efficiency of conversion of rest mass into heat energy  
of any element of matter transversing the discs from their outer edge located
on the outer marginally stable  
orbit to their inner edge located on the inner marginally stable orbit is
given by 
\begin{equation}             
\eta \equiv E_{\mathrm{ms(o)}}-E_{\mathrm{ms(i)}}.
\end{equation}
For Keplerian discs co-rotating extreme KdS black holes, the accretion
efficiency 
reaches maximum value of $\eta \sim 0.43$ for the pure Kerr case ($y=0$) and
tends  
to zero for $y \to y_{\rm c(KdS)}\doteq 0.059$, the maximum value of $y$
admitting black holes\footnote{Notice that for the plus-family discs orbiting
  Kerr naked singularities with $a \sim 1$, the efficiency $\eta \sim 1.57$,
  exceeding strongly the annihilation efficiency. This is caused by strong
  discontinuity in properties of the plus-family orbits for extreme black
  holes and 
  naked singularities with $a \to 1$. Conversion of a naked singularity into
  an extreme black hole leads to an abrupt instability of the innermost parts
  of the plus-family discs that can have strong observational
  consequences.\cite{Stu-Sla:2004:PHYSR4:}}.

\section{Thick discs}

Basic properties of thick discs are determined by equilibrium configurations
of perfect fluid. Stress-energy tensor of perfect fluid is given by 
\begin{equation}
     T^\mu_{\hphantom{\mu}\nu} = (p+\epsilon) U^\mu U_\nu + p\,\delta^\mu_\nu
\label{eq1}
\end{equation}
where $\epsilon$  and~$p$ denote total energy density and~pressure of~the
fluid, $U^{\mu}$ is its four velocity. 
We shall consider test perfect fluid rotating in~the $\phi$ direction, i.e.,
$U^{\mu} = \left( U^{t}, U^{\phi}, 0, 0\right)$. The rotating fluid can be
characterized by the 
vector fields of the angular velocity $\Omega \left( r, \theta \right)$
and~the angular momentum density 
$\ell \left( r, \theta \right)$, defined by
\begin{equation}
     \Omega = \frac{U^{\phi}}{U^{t}}, \qquad \ell = - \frac{U_{\phi}}{U_{t}}.
\label{eq2}
\end{equation}
The vector fields are related by the metric coefficients of~the KdS spacetime 
\begin{equation}
     \Omega = - \frac{g_{t\phi} + \ell g_{tt}}{g_{\phi\phi} + \ell g_{t\phi}}.
\label{eq3}
\end{equation}

Projecting the energy-momentum conservation law
$T^{\mu\nu}_{\hphantom{\mu\nu};\nu} = 0$ onto the hypersurface orthogonal to
the four velocity $U^\mu$ by the projection tensor $h_{\mu\nu} = g_{\mu\nu} +
U_\mu U_\nu$, we obtain the relativistic Euler equation in the form
\begin{equation}
  \frac{\partial_{\mu} p}{p+\epsilon} =
    -\partial_{\mu} (\ln U_t) +
    \frac{\Omega\,\partial_{\mu} \ell}{1- \Omega \ell},          \label{eqv6}
\end{equation}
where
\begin{equation}
  (U_t)^2 =
    \frac{g^2_{t\phi} - g_{tt}\,g_{\phi\phi}}%
         {g_{\phi\phi} + 2 \ell g_{t\phi} + \ell^2 g_{tt}}.      \label{eqv7}
\end{equation}

For barytropic perfect fluid, i.e., the fluid with an equation of state
$p=p(\epsilon)$, the solution of the relativistic Euler equation can be given
by Boyer's condition determining the surfaces of constant pressure through the
``equipotential surfaces'' of the potential $W(r,\theta)$ by the
relations\cite{Abr-Jar-Sik:1978:ASTRA:}
\begin{eqnarray}
  &&\int_0^p \frac{\d p}{p+ \epsilon} = W_\mathrm{in} - W =
    \ln (U_t)_\mathrm{in} - \ln (U_t) +
    \int_{\ell_\mathrm{in}}^{\ell}
    \frac{\Omega\,\d\ell}{1- \Omega \ell};                       \label{eqv9}
\end{eqnarray}
the subscript ``in'' refers to the inner edge of the disc. 

The equipotential surfaces are determined by the condition
\begin{equation}
  W(r,\theta) = \mathrm{const},
\label{eqv10} 
\end{equation}
and in a given spacetime can be found from Eq.~(\ref{eqv9}), if a rotation
law $\Omega = \Omega(\ell)$ is given.
Equilibrium configurations of~test perfect fluid are determined by~the
equipotential surfaces which can be closed or open.
Moreover, there is a special class of critical, self-crossing surfaces (with a
cusp), which can be either closed or open. The closed equipotential surfaces
determine stationary toroidal configurations.  The fluid can fill any closed
surface -- at the surface of the equilibrium configuration pressure vanishes,
but its gradient is non-zero.\cite{Koz-Jar-Abr:1978:ASTRA:} On the other
hand, the open equipotential surfaces are important in dynamical situations,
{\it e.g.}, in modeling of jets.\cite{LyB:1969:NATURE:,Bla:1987:300YoG:} The
critical, self-crossing closed equipotential surfaces $W_\mathrm{cusp}$ are
important in the theory of thick accretion discs, because accretion onto the
black hole through the cusp of the equipotential surface, located in the
equatorial plane, is possible due to a little overcoming of the critical
equipotential surface by the surface of the disc (Paczy\'nski mechanism). 
Accretion is thus driven by a violation of the hydrostatic equilibrium,
rather than by viscosity of the accreting matter.\cite{Koz-Jar-Abr:1978:ASTRA:}

It is well known that all characteristic properties of the equipotential
surfaces for a general rotation law are reflected by the equipotential
surfaces of the simplest configurations with uniform distribution of the
angular momentum density $\ell$, see Ref.~\refcite{Jar-Abr-Pac:1980:ACTAS:}.
Moreover, these
configurations are very important astrophysically, because they are marginally
stable.\cite{Seg:1975:ASTRJ2:}
Under the condition
\begin{equation}
  \ell(r,\theta) = \mathrm{const},
\label{eqv11} 
\end{equation}
a simple relation for the equipotential
surfaces follows from Eq.~(\ref{eqv9}):
\begin{equation}
  W(r, \theta) = \ln U_t (r, \theta).                           \label{eqv12}
\end{equation}

The equipotential surfaces $\theta = \theta(r)$ are given by the relation
\begin{equation}
  \oder{\theta}{r} =
    -\frac{\partial p/\partial r}{\partial p/\partial \theta},  \label{eqv13}
\end{equation}
which for the configurations with $\ell = \mathrm{const}$ reduces to
\begin{equation}
  \oder{\theta}{r} =
    -\frac{\partial U_t/\partial r}{\partial U_t/\partial\theta}.\label{eqv14}
\end{equation}
In~the KdS spacetimes there is
\begin{equation}
     W \left(r, \theta \right)    = \ln \left\{ \frac{\rho}{I} \cdot
     \frac{\Delta^{1/2}_r \Delta^{1/2}_{\theta} 
                                        \sin \theta}{\left[\Delta_{\theta}
     \sin^{2}\theta 
                                        \left(r^2 + a^2 - a\ell \right)^{2} -
     \Delta_{r} 
                                        \left(\ell - a\sin^{2}\theta
     \right)^{2}\right]^{1/2}}\right\}. 
\end{equation}
The best insight into the $\ell = \mbox{const}$ configurations is given by
properties of $W\left(r, \theta\right)$ in the equatorial plane ($\theta =
\pi/2$). The reality conditions of $W\left(r, \theta = \pi/2\right)$ imply
\begin{equation}
     \ell_{\rm ph-} < \ell < \ell_{\rm ph+},
\label{eq5}
\end{equation}
where the functions $\ell_{\rm ph \pm} \left(r; a, y\right)$, given by 
\begin{equation}
     \ell_{\rm ph\pm}(r;a,y) = a + \frac{r^2}{a\pm\sqrt{\Delta_r}},
                                                              \label{eq7}
\end{equation}
determine the photon geodesic
motion.\cite{Stu-Hle:2000:CLAQG:,Stu-Sla:2004:PHYSR4:}

\begin{figure}
\includegraphics[width=1\hsize]{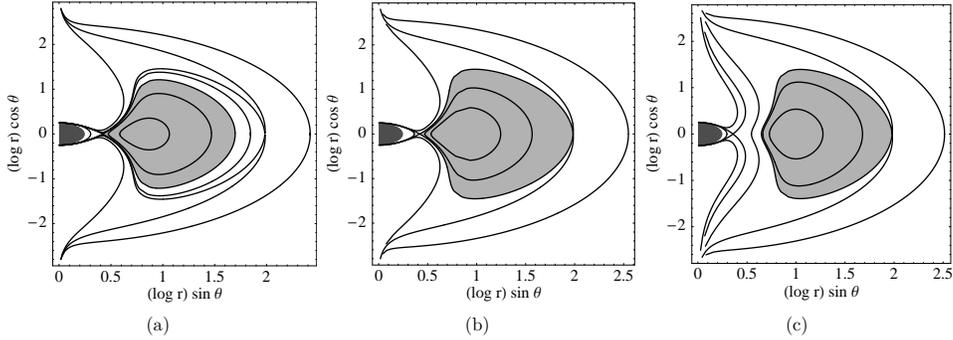}
\caption{Typical behaviour of equipotential surfaces
(meridional sections) in the KdS black-hole spacetimes. Light gray region
contains closed equipotential surfaces. The last closed surface is
self-crossing in the cusp(s). Possible toroidal configurations correspond to: 
(a) accretion discs, (b) marginally bound accretion discs and (c) excretion
discs.}
\label{f3}
\end{figure}

Condition for the local extrema of the potential $W\left(r, \theta =
\pi/2\right)$ is identical with the condition 
of~vanishing of~the pressure gradient $\left(\lpder{U_t}{r} = 0 =
\lpder{U_t}{\theta}\right)$. 
Since in~the equatorial plane there is $\lpder{U_t}{\theta} = 0$, independently
of $\ell = \mbox{const}$, 
the only relevant condition is $\lpder{U_t}{r} = 0$,
which implies the relation
\begin{equation}
      \ell=\ell_{\rm K \pm}(r;a,y)
\end{equation}
with $\ell_{\rm K \pm}$ being the angular momentum density of the geodetical
Keplerian orbits
\begin{equation}
\ell_{\rm K \pm} \left(r; a, y\right) \equiv \pm
  \frac{(r^{2}+a^{2})(1-yr^{3})^{1/2}\mp ar^{1/2}[2+yr(r^{2}+a^{2})]}%
       {r^{3/2}[1-y(r^{2}+a^{2})]-2r^{1/2}\pm a(1-yr^{3})^{1/2}}.
                                                               \label{eq12}
\end{equation}
The closed equipotential surfaces, and surfaces with a~cusp allowing the
outflow of matter from~the disc, are permitted in~those parts of~the functions
$\ell_{\rm K \pm} \left(r; a, y\right)$ enabling the existence of~stable
circular geodesics corresponding to~the centre of~the equilibrium
configurations. Stationary toroidal configurations exist if 
$\ell\in (\ell_{\rm ms(i)},\ell_{\rm ms(o)})$.
We can distinguish three kinds of discs (Fig.~\ref{f3}):
\begin{description}
\item [accretion discs:] $\ell\in (\ell_{\rm ms(i)},\ell_{\rm mb})$; the last
  closed 
  surface is self-crossing in the inner cusp, another critical surface
  self-crossing in the outer cusp is open. 
\item [marginally bound accretion discs:] $\ell=\ell_{\rm mb}$;  the last
  closed 
  surface is self-crossing in both the inner and the outer cusp.
\item [excretion discs:] $\ell\in (\ell_{\rm mb},\ell_{\rm ms(o)})$; the
last closed surface is self-crossing in the outer cusp, another critical
surface self-crossing in the inner cusp is open.
\end{description}

%%%%%%%%%%%%%%%%%%%%%%%%%%%% figure 3 %%%%%%%%%%%%%%%%%%%%%%%%%%%%%%%%%%%%%%
%%%%%%%%%%%%%%%%%%%%%%%%%%%%%%%%%%%%%%%%%%%%%%%%%%%%%%%%%%%%%%%%%%%%%%%%%%%%

\section{Conclusions}

For astrophysically relevant black holes ($M < 10^{12}M_{\odot}$)
and the observed RRCC~(\ref{E3}), the 
cosmological parameter is so small ($y < 10^{-22}$) that both co-rotating 
and counter-rotating discs can exist around KdS black holes. The efficiency
of the accretion process is then extremely close to the values relevant for 
Kerr black holes. The efficiency is strongest for thin, Keplerian discs
orbiting extreme black holes. It is suppressed for $a$ descending and/or for
$\ell = \mbox{const}$ growing from $\ell_{\rm ms(i)}$ up to $\ell_{\rm mb}$. 
Notice that the co-rotating toroidal discs are steeper and more extended 
than the counter-rotating discs.

\begin{figure}
\centering
\includegraphics[width=.9\hsize]{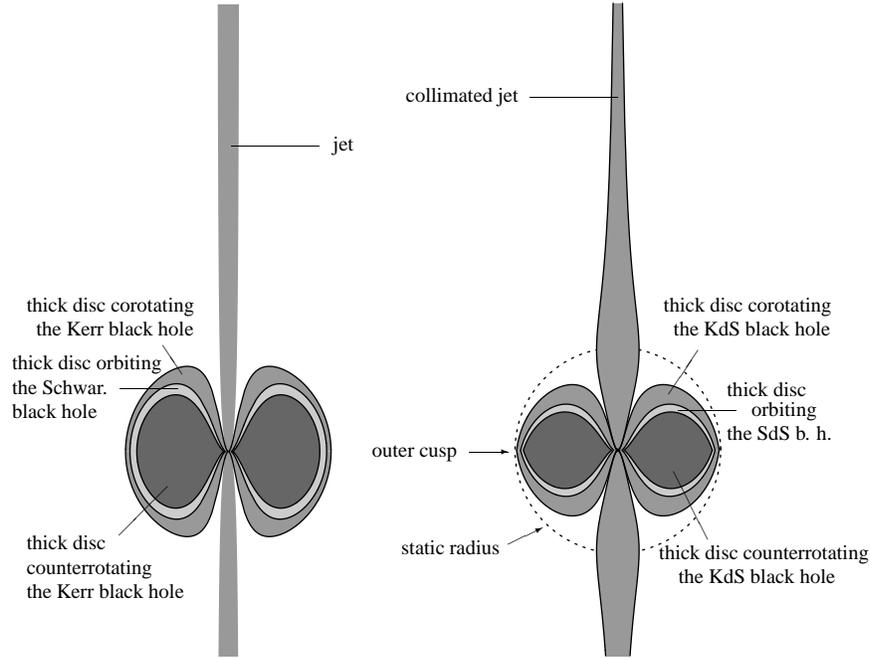}
\caption{Shapes of thick discs and collimation of jets due to a cosmic
  repulsion. The effect of collimation is relevant near the static radius and
  further. Left picture depicts thick accretion discs orbiting the Kerr
  black hole ($y=0,\ a^2=0.99;\ \ell\approx\ell_{\rm mb}$) and the
  Schwarzschild black hole ($y=0,\ a=0;\ \ell\approx\ell_{\rm mb}$),  
right picture depicts thick marginally bound accretion discs orbiting the KdS
  black hole ($y=10^{-6},\ a^2=0.99;\ \ell=\ell_{\rm mb}$) and the SdS black
  hole ($y=10^{-6},\ a=0;\ \ell=\ell_{\rm mb}$).}
\label{f4}
\end{figure}

The crucial effects caused by the RRCC are illustrated in (Fig.~\ref{f4}). 
\begin{itemize}
\item 
The outer edge of the discs. The presence of an outer cusp of toroidal discs
nearby the 
static radius enables outflow of mass and angular momentum from the discs
due to a violation of mechanical equilibrium. Recall that such an outflow is
impossible from discs around isolated black holes in asymptotically flat
spacetimes.\cite{Koz-Jar-Abr:1978:ASTRA:}
\item 
Strong collimation effect on jets escaping along the rotational axis of
toroidal discs 
indicated by open equipotential surfaces that are narrowing strongly after
crossing the static radius.
\end{itemize}

%%%%%%%%%%%%%%%%%%%%%%%%%%% figure 4 %%%%%%%%%%%%%%%%%%%%%%%%%%%%%%%%%%%%%%%
%%%%%%%%%%%%%%%%%%%%%%%%%%%%%%%%%%%%%%%%%%%%%%%%%%%%%%%%%%%%%%%%%%%%%%%%%%%%%

We can give to our results proper astrophysical relevance by presenting
numerical
estimates for observationally established current value of the
RRCC\footnote{For more detailed information in the case of thick discs 
around Schwarzschild--de Sitter black holes see
Ref.~\refcite{Stu-Sla-Hle:2000:ASTRA:}, where the
estimates for primordial black holes in the early universe with a
repulsive cosmological constant related to a hypothetical vacuum
energy density connected with the electroweak symmetry breaking or the
quark confinement are presented.}. 
Having the value of $\Lambda_{0} \approx 1.3 \times 10^{-56}{\rm cm}^{-2}$, 
we can determine the mass parameter
of the spacetime corresponding to any value of $y$, parameters of the
equatorial circular geodesics and basic characteristics of both the thin and
thick accretion discs (Table~\ref{t1}). Outer edge of the marginally bound
thick accretion disc is
determined by the outer marginally bound circular orbit which is located
very close to, and for presented values of $y$ almost at
the static radius of a given spacetime. 

%%%%%%%%%%% table 1 %%%%%%%%%%%%%%%%%%%%%%%%%%%%%%%%%%%%%%%%%%%%%%%%%%%%%%%%%%%
\begin{table}
\centering
\tbl{Mass parameter, the static radius and radius of the outer 
marginally stable orbit in extreme KdS black-hole
spacetimes are given for the RRCC
indicated by recent cosmological observations.}
{\begin{tabular}{lcccccc}
\toprule
$y$ & $10^{-44}$ & $10^{-34}$ & $10^{-30}$ & $10^{-28}$ & $10^{-26}$ &
$10^{-22}$ \\ 
$M/M_{\odot}$ & 10 & $10^6$ & $10^8$ & $10^9$ & $10^{10}$ & $10^{12}$ \\
$r_{\mathrm{s}}/\mbox{[kpc]}$ & 0.2 & 11 & 50 & 110 & 230 & 1100 \\
$r_{\mathrm{ms}}/\mbox{[kpc]}$ & 0.15 & 6.7 & 31 & 67 & 150 & 670 \\
\botrule
\end{tabular}}
\label{t1}
\end{table}
%%%%%%%%%%%%%%%%%%%%%%%%%%%%%%%%%%%%%%%%%%%%%%%%%%%%%%%%%%%%%%%%%%%%%%%%%%%%%%%

It is well known\cite{Car-Ost:1996:ModAst:} that 
dimensions of accretion discs around stellar-mass black holes ($M
\sim 10 M_\odot$) in binary systems are typically $10^{-3}$\,pc, dimensions of
large galaxies with central black-hole mass $M \sim 10^8 M_\odot$, of
both spiral and elliptical type, are in the interval 50--100 kpc, and
the extremely large elliptical galaxies of cD type with central
black-hole mass $M \sim 3\times 10^9 M_\odot$ extend up to 1
Mpc. Therefore, we can conclude that the influence of the RRCC 
is quite negligible in the accretion discs in
binary systems of stellar-mass black holes as the static radius
exceeds in many orders dimension of the binary systems. But it can be
relevant for accretion discs in galaxies with large active nuclei as
the static radius puts limit on the extension of the discs well inside 
the galaxies. Moreover, the agreement (up to one order) of the
dimension of the static radius related to the mass parameter of
central black holes at nuclei of large galaxies
with extension of such galaxies suggests that the RRCC could play 
an important role in formation and evolution of such
galaxies. Of course, the first step in confirming such a suggestion is
modelling of the influence of the RRCC on self-gravitating accretion discs.

\section*{Acknowledgments}

The present work was supported by the Czech grant MSM 4781305903 and by the 
Committee for Collaboration of Czech Republic with CERN. The author 
would like to acknowledge Drs Stanislav Hled\'{\i}k and Petr Slan\'{y} for
collaboration 
and the excellent working conditions at the CERN's Theory Division and 
SISSA's Astrophysics Sector, respectively, where part of the work was realized.

\end{document}